\documentclass{article} 
\usepackage{graphicx}
\usepackage{dcolumn}
\usepackage{epstopdf}
\usepackage{amsmath}
\usepackage{amssymb}
\usepackage{wasysym}
\usepackage{color}
\usepackage{natbib}

\title{Marangoni spreading due to a localized alcohol supply on a thin water film}
\author{Jos\'e Federico Hern\'andez-S\'anchez, Antonin Eddi and J.H.~Snoeijer\\
\\ Physics of Fluids Group, Faculty of Science and Technology,\\ J.M. Burgers Center for Fluid Dynamics, University of Twente,\\ P.O. Box 217, 7500 AE  Enschede, The Netherlands}
\date{\today}

\begin{document}
\maketitle
\begin{abstract}
Bringing the interfaces of two miscible fluids into contact naturally generates strong gradients in surface tension. Here we investigate such a Marangoni-driven flow by continuously supplying isopropyl alcohol (IPA) on a film of water, using micron-sized droplets of IPA-water mixtures. These droplets create a localized depression in surface tension that leads to the opening of a circular and thin region in the water film. At the edge of the thin region, there is a rim growing and collecting the water of the film. We find that the spreading radius scales as $r \sim t^{1/2}$. This result can be explained from a balance between Marangoni and viscous stresses, assuming that the gradients in surface tension are smoothened out over the entire size of the circular opening. We derive a scaling law that accurately predicts the influence of the IPA flux as well as the thickness of the thin film at the interior of the spreading front.
\end{abstract}

\section{Introduction}

Surface tension gradients appear when liquids with different surface tensions or with different surfactants concentrations are put into contact, leading to Marangoni flows. A lot of work focussed on the radial spreading induced by a localized gradient of surfactant concentration \citep{Borgas1988, Gaver1990, Jensen1994, Matar1999, Hamraoui2004, Matar2004, Fallest2010}. For insoluble surfactants, the concentration only affects the local surface tension leaving the bulk properties unaffected ({\it e.g.} viscosity and density), which allows to use a full theoretical approach. \cite{Borgas1988} investigated the process experimentally and proposed the physical quantities involved in the process, while \cite{Gaver1990} used the lubrication equation to describe the phenomena in thin films. Later, \cite{Jensen1994} investigated similarity solutions for the spreading of a surfactant on a thin film. For an axisymmetric spreading of a fixed amount of surfactant and a source, the scaling laws for the radial position with time are $r\sim t^{1/4}$ and $r\sim t^{1/2}$, respectively. \cite{Shearer2011} extended this result giving similarity solutions and numeric simulations for a finite amount of surfactant. Experimentally, the $r\sim t^{1/4}$ was confirmed by \cite{Fallest2010}, in a study that also measured the concentration profiles. Some of these results have been extended to spreading on thick layers, for which $r\sim t^{3/8}$ \citep{Jensen1995}.

All the previous results were obtained for insoluble surfactants, but some studies focused on the influence of solubility. \cite{Afsar_PART1_2003} and \cite{Afsar_PART2_2003} performed a series of experiments investigating the effect of solubility in the Marangoni drop spreading and found that the spreading exponent is $r\sim t^{1/4}$. More recently, \cite{Lee2009} determined the initial effect of the drop geometry and confirmed this result. A review of the Marangoni spreading can be found in \cite{matar2009dynamics}. By contrast, Marangoni flows induced by transport of two miscible liquids of different liquids of different surface tensions has received less attention. A famous example arises for the tears of wine \citep{Hosoi2001}, which are complicated by evaporation dynamics. Another path is to study the Marangoni forces involved during the drying of surfaces. The industrial technique called Marangoni drying was developed in the nineties and was experimentally studied by \cite{Leenaars1990} and \cite{Marra1991}. Theoretical development was achieved by \cite{OBrien1993} and \cite{Matar2001} by using the lubrication approximation to find the height profiles for the interaction of a thin liquid layer with a volatile liquid. Using a very different geometry than the radial spreading of surfactant, \cite{Karpitschka2010} studied the effect of Marangoni forces in delaying the coalescence of miscible sessile drops. They found that the bridge joining the drops displaces at a constant velocity. 

In this paper, we experimentally investigate the spreading dynamics due a continuous supply of a low surface tension liquid, on a water thin film. In section 2 we describe the experimental setup, the measurements and the calibrations required to achieve well controlled quantitative experiments. In section 3 we present measurements of the radius of the rim as a function of time for different conditions and propose a scaling argument, describing the process as the dynamic balance between the Marangoni and viscous stresses. Here, we explore the influence of different parameters ({\it i.e.}, initial thickness of the water layer and liquid flux and surface tension) and compare it with the scaling theory by collapsing the data. In section 4, we present our conclusions and discussion and in section 5 we give an appendix with some further experimental details.

\section{Experimental setup}

\begin{figure}
		\centerline{\includegraphics[width=1.\columnwidth]{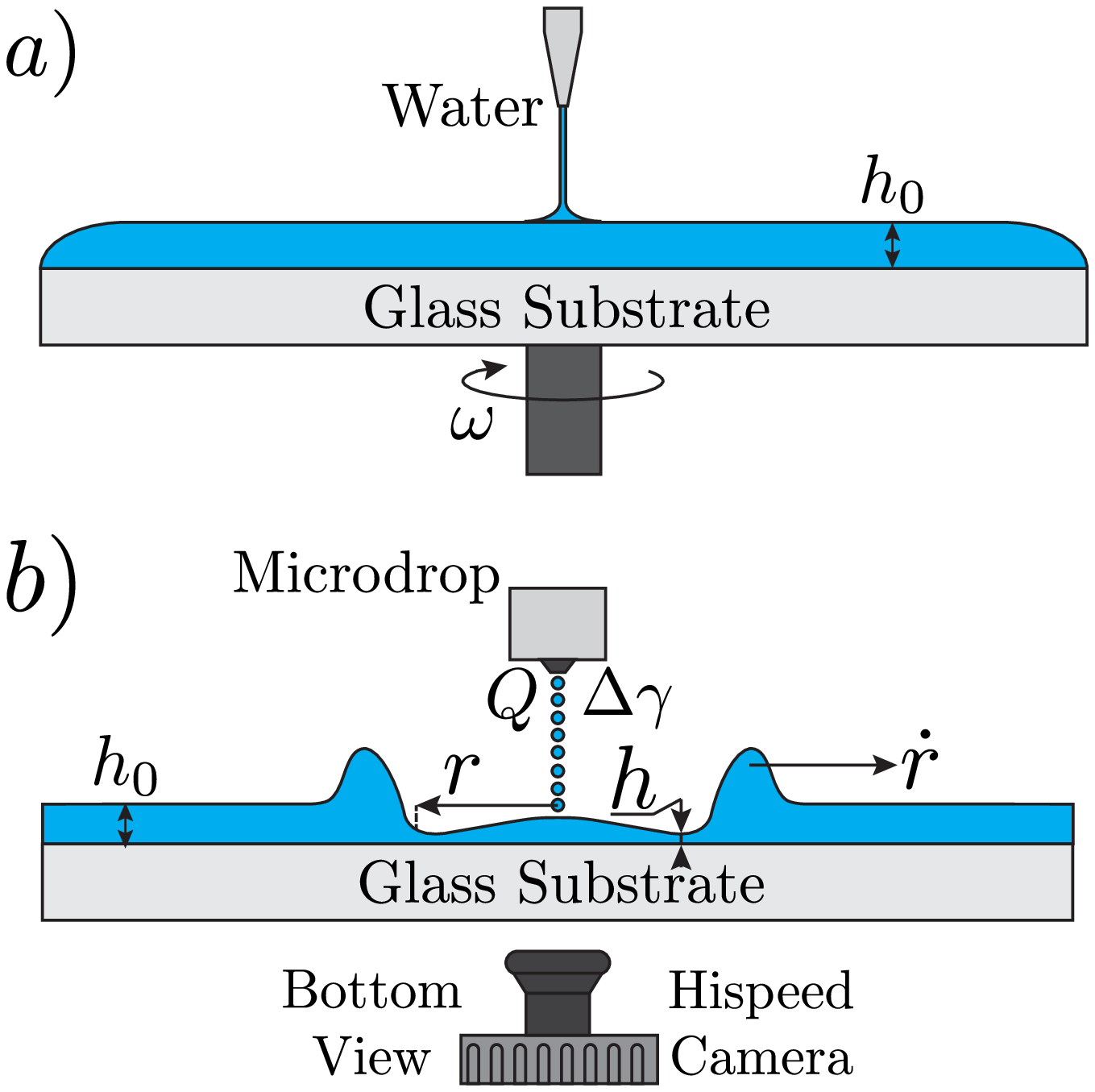}}
\caption{\label{fig1}($a$) Sketch of the spin coater setup, depositing a uniform film of water of thickness $h_0$. ($b$) Sketch of the radial spreading dynamics viewed as cross-section. After placing the substrate on the inverted microscope, a color high speed camera is used to record the process. A microdrop nozzle shoots droplets ($d \sim 50\mu m$) with controlled frequency of an IPA-water mixture on the water layer. The flux $Q$ of IPA-water induces a surface tension difference $\Delta \gamma$. ($c-f$) Bottom view of the typical experiment image from the inverted microscope at times $t=5ms$, $t=37ms$, $t=97ms$, $t=157ms$. The experimental conditions are $h_0 = 14\mu s$, $Q=157nL/s$, $1\% IPA (Vol.)$. The interferometry patterns are the result of a small thickness in the region behind the outward moving front.}
\end{figure}

Here we describe the experimental setup, measurement procedure and necessary calibrations. Two steps were followed for each experiment. First, a water layer of uniform thickness is deposited on a hydrophilic substrate using a spin coater [Fig.~\ref{fig1}(a)]. The thickness is varied from 8 to 70 microns, where the thickness was determined using a high-resolution spectrometer (Ocean optics HR4000). The substrate consists of a silica glass slide ($71\times71m$m), which is made hydrophilic using the cleaning procedure described in the Appendix. This cleaning step is critical to achieve reproducible results. Second, the substrate with the film is placed on the inverted microscope (Zeiss Axiovert 25), with a high speed color camera (Photron SA2) recording the bottom view [Fig.~\ref{fig1}(b)]. A Marangoni flow is created by a continuous supply of micron-sized drops of an IPA-water mixture, deposited on top of the water film. The IPA-water mixture has a lower surface tension than water, giving rise to localized depression of surface tension and hence Marangoni forces. As seen in Fig.~\ref{fig1}(c-f), these forces induce a radially outward flow in the form of a circular traveling front, leaving behind a liquid layer much thinner than the initial thickness. The small drops that are deposited at the center of the images have negligible inertia, ensuring that this radial spreading are solely driven by Marangoni forces. While the growth of the circular front is reminiscent of the classical ``dewetting hole" (dry circular patches on partially wetting surfaces \citet{DeGennes1992}), a key difference is that here the substrate is perfectly wetting and a macroscopic liquid film remains in the interior.

\begin{figure}
	\includegraphics[width=.99\columnwidth]{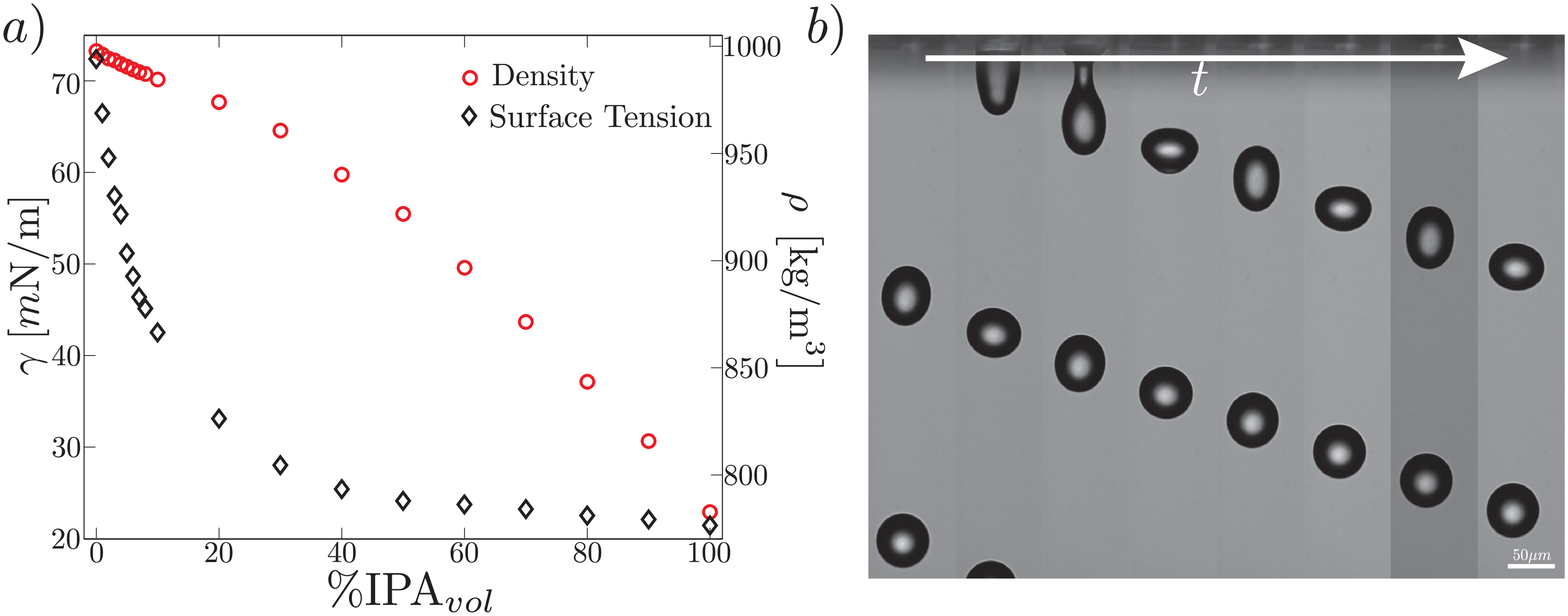}
\caption{\label{fig5} (a) Calibration of the density ($\bigcirc$, right vertical axis) and the surface tension($\Diamond$, left vertical axis) as a function of IPA concentration. (b) Visualization of the generation of microdrops, obtained using iLIF technique developed by \cite{vanderBos2011}. From these type of images, we could determine the range over which the monodisperse train of droplets was generated, and determine the corresponding flow rates $Q$.}
\end{figure}

The goal of the experiment is to reveal the dynamics of the spreading radius $r(t)$, as a function of the control parameters of the experiment. Through image analysis, we measure the radius of the front in the recordings. The concentric rings in the inner region appear due the constructive light interference. Using the color interferometry technique reported by \cite{Roeland2012}, the fringes provide a measurement of the film thickness in the thinnest region. We performed experiments by systematically varying the initial film thickness $h_0$, the flow rate of the IPA-water drops $Q$, as well as the composition of the IPA-water mixture. By changing the composition, we vary the surface tension and the density as has been calibrated in Fig.~\ref{fig5}(a). The density and the surface tension were measured using the density meter (DM A35 from Anton Paar) and the pendant drop method. These calibrations show excellent agreement with \cite{ViscoProp01}, who previously measured surface tension and viscosity measurements of IPA-water mixtures. We vary the flow rate $Q$ by the frequency at which the microdrops are deposited. The exact flow rate is calibrated by measuring the droplet size for each frequency, using the technique developed by \cite{vanderBos2011}, from typical images shown in Figure.~\ref{fig5}(b).

\section{Results and scaling laws}

\begin{figure}
	\centerline{\includegraphics[width=0.5\columnwidth]{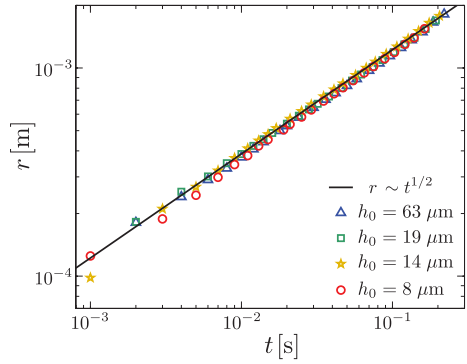}}
\caption{\label{fig2} Radius of the rim opening as a function of time for different thickesses of the water layer film $h_0$. The symbols ($\bigtriangleup$), ($\Box $), ($\star$) and ($\bigcirc$) respectively correspond to initial water layer thickess $h_0=63\mu$m, $h_0=19\mu$m,  $h_0=14\mu$m and $h_0=8\mu$m), respectively. The liquid flow to the surface is $ Q=150.7n$L/s and the isopropanol concentration is $1\%$ which leads to a surface tension difference $\Delta\gamma=5.56m$N/m. The initial time is taken at the instant of contact observed in the recordings. The dynamics of the front is found to be independent of the initial water layer thickness $h_0$. Each curve is the average of three experiments under the same conditions.}
\end{figure}

We start by considering a 1\% (in volume) concentration IPA-water mixture that is deposited at a flow rate $Q=150.7n$L/s, onto films of varying initial thickness $h_0$. The results are presented in Fig.~\ref{fig2}, showing the front position as a function of time on a log-log plot. We observe that the radius scales as $r\sim t^{1/2}$. Below we will show that this scaling law is very robust and observed for all experiments. From the measurements in Fig.~\ref{fig2} we find no influence of the initial thickness. 

\begin{figure}
		\centerline{\includegraphics[width=1.\columnwidth]{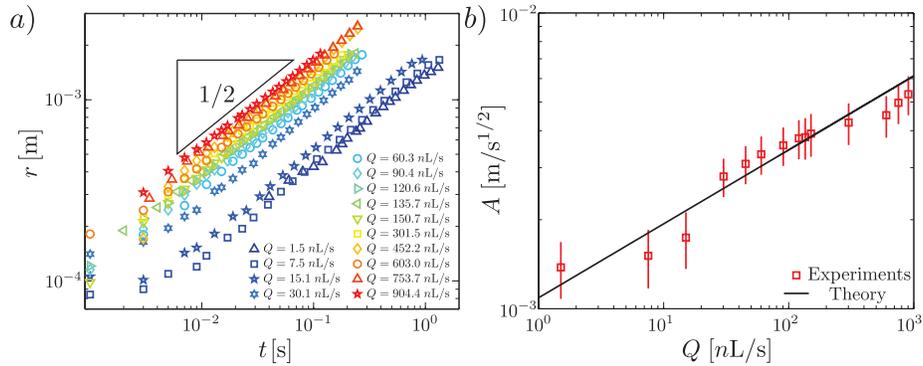}}
\caption{\label{fig3} (a) Measurement of the radius of the front as a function of time for different rates of IPA-water supply. The flow rate was varied from $ Q=1.5n$L/s to $904n$L/s. The spreading dynamics follows $r(t) = A t^{1/2}$, where the prefactor $A$ increases with $Q$. (b) The fitted values of $A$ as a function of the flow rate $Q$. The solid line correponds to the predicted scaling law $A\sim {Q}^{1/4} $, with a prefactor 0.7. The initial thickness in these experiments is $h_0=14\mu$m.}
\end{figure}

Next we consider the influence of the flux $Q$, while keeping the IPA concentration constant at 1\%. The variation in the flux $Q$ spans nearly three orders of magnitude from $1.5n$L/s to $904.4n$L/s. The initial thickness in these experiments is kept constant at $h_0=14\mu$m. In Fig.~\ref{fig3}(a), we show the radius as a function of time for different $Q$. This result confirms that the dynamics is indeed very robust and follows $r \sim t^{1/2}$, independent of the rate at which IPA is supplied. However, we do observe a clear influence of $Q$ on the dynamics: increasing the flow rates leads to a faster Marangoni flow. To quantify this result, we fitted the curves in Fig.~\ref{fig3}(a) by

\begin{equation}
	r=A\;t^{1/2},
	\label{eq:onehalftime}
\end{equation}
and determined the prefactor $A$ as a function of $Q$. As shown in Fig.~\ref{fig3}(b), the data are consistent with a power-law, $A\sim Q^{n}$. Fitting the value of the exponent, we obtain $n=0.23\pm0.02$. Below we argue that this corresponds to a scaling exponent $n=1/4$, which is shown as the solid black line.

These observations can be explained from a balance between Marangoni-driving and viscous dissipation. Here we provide an scaling analysis using similar ideas as \cite{Jensen1994}. The front is driven outwards by a Marangoni stress,

\begin{equation}
	\tau_{M}=\frac{d \gamma}{d r} \sim\frac{\Delta \gamma}{r},
	\label{eq:mstress}
\end{equation}
where $\Delta \gamma$ is the surface tension difference between the two liquids. By scaling the $r$-derivative as $1/r$, we assume that the surface tension difference spreads out over the entire radius of the inner circle. The Maragoni-tress $\tau_M$ is opposed by a viscous stress $\tau_\eta$, which for thin film flows reads

\begin{equation}
	\tau_{\eta}=\eta \frac{d v_r}{d z} \sim \eta \frac{ \dot{r}}{h}.
	\label{eq:visstress}
\end{equation}
where $\eta$ is the viscosity, $\dot{r}=dr/dt$ is the velocity of the front and, $h$ is the characteristic thickness of the film. Balancing $\tau_M \sim \tau_\eta$ we find

\begin{equation}
	\frac{\Delta \gamma}{r}  \sim \eta \frac{\dot{r}}{h}.
	\label{eq:balan}
\end{equation}

Crucially, we have seen that the initial thickness of the film $h_0$ is irrelevant for the dynamics. Hence, we assume that the characteristic thickness $h$ appearing in (\ref{eq:balan}) corresponds to the thin film at the interior of the circle, the thickness of which is an unknown dynamical variable as well. To close the problem we use mass conservation: we consider that the flux of incoming drops $Q$ equals the flux displaced by the movement of the front. The latter scales like the front velocity $\dot{r}$ times the area associated to the perimeter, $2\pi r h$, which in terms of scaling gives

\begin{equation}
	 Q  \sim  r h \dot{r}.
	\label{eq:flux1}	
\end{equation}
Combined with Eq.~(\ref{eq:balan}), we can eliminate $h$ and obtain, 

\begin{equation}
	Q \sim  \frac{\eta}{\Delta \gamma}\;  r^{2}  \; {\dot{r}}^{2},
	\label{eq:flux2}
\end{equation}
or equivalently,

\begin{equation}
	r \sim \left(\frac{ Q\;\Delta\gamma}{\eta}\right)^{1/4} t^{1/2}, \quad \Longrightarrow \quad A = k  \left(\frac{ Q\;\Delta\gamma}{\eta}\right)^{1/4}.
	\label{eq:rdt}
\end{equation}
Here we introduced $k$ as a numerical prefactor that we expect to be of order unity. Tracing back the steps of the similarity analysis for insoluble surfactants by \cite{Jensen1994}, our results correspond to the particular case of a source constantly adding mass with time in an axisymmetric geometry ($\alpha=0$, $\delta=1/2$, $p=1$ and $n=1$).

These scaling arguments indeed explain the two key experimental observations: $r \sim t^{1/2}$ and $A \sim Q^{1/4}$. In addition, Eq.~(\ref{eq:rdt}) gives a prediction for $A$ in terms of the known parameters $\Delta\gamma = 5.56\;m$N/m, and $\eta=0.95m\mbox{Pa}\cdot\mbox{s}$. We use the dataset in Fig.~\ref{fig3}(b) to determine the numerical prefactor $k$. The black solid line indeed corresponds to a value of order unity, namely $k=0.7$. 

\begin{figure}[h]
	\centerline{\includegraphics[width=1.\columnwidth]{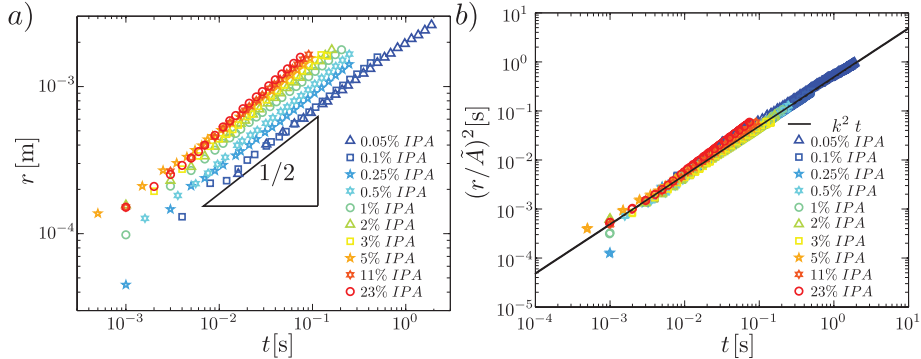}}
\caption{\label{fig4}(a) Measurement of the front radius $r$ as a function of time for different experiments using different isopropanol concentrations. The initial time is taken at the instant of contact observed in the recordings. The isopropanol concentrations were varied from $0.05\%$ to $23\%$ which lead to surface tension differences from $\Delta \gamma = 0.3mN/m$ to $40.61mN/m$. (b) Collapsed curve for the showed isopropanol concentrations. The curves were normalized using $A = \left(Q \Delta \gamma / \eta \right)^{1/4}$.  The initial thickness in these experiments is $h_0=14\mu$m.}
\end{figure}

To further test the scaling theory, we perform a set of experiments changing the isopropanol concentration in the IPA-water mixture from $0.05\%$ to $23\%$. This changes both the surface tension and the viscosity of the IPA-water mixture, which in practice means that we have to recalibrate the settings of the microdrop for given frequency and isopropanol concentration. Therefore, by changing the IPA concentration, we vary $Q$, $\eta$ and $\Delta\gamma$ at the same time. Yet, all values have been calibrated independently, allowing for a quantitative comparison to the theory. In Fig.~\ref{fig4}(a) we present the results for the different IPA-water mixtures, corresponding to a large range of surface tension differences, from $\Delta\gamma=0.3\mbox{m}N/m$ to $40\mbox{m}N/m$. Once more, we find $r \sim t^{1/2}$ and we can directly test the prediction for $A$. To this end, we try to collapse the data by plotting $(r/\tilde{A})^2$, where $\tilde{A}=(Q\;\Delta\gamma/\eta)^{1/4}$ should account for the parametric dependence. Indeed, the data nicely collapse as shown in Fig.~\ref{fig4}(b), without adjustable parameter. The solid line corresponds to the theory with $k=0.7$. We conclude that the scaling theory very well describes our experimental observations.

Finally, one can solve the thickness of the thin film $h$, by combining Eq.~(\ref{eq:balan}) and~(\ref{eq:rdt}). The various scaling laws are such that the thickness is independent of time,
\begin{equation}
	h\sim  \left(\frac{\eta\; Q}{\Delta\gamma}\right)^{1/2},
	\label{eq:hmin}
\end{equation}
and which is expressed in terms of known quantities. For the experiment in Fig~\ref{fig1}(c-f), these are $\Delta\gamma = 5.56\;m$N/m, $Q=150.7n$L/s and $\eta=0.95m$Pa$\cdot$s, such that $h\sim 5.1\mu$m. This estimate gives the characteristic thickness of the region at the interior of the circular region. Owing to the fringes observed just behind the moving front [see e.g. Fig.~\ref{fig1}(f)], we can locally determine the film thickness using the interferometry technique developed by \cite{Roeland2012}. Indeed, we obtain a thickness that during the course of the experiment is approximately constant, within a range of $\pm250n$m, and that is determined $h \sim 1.4\mu$m. It is a bit smaller than the prediction by the scaling law (\ref{eq:hmin}), though still of comparable magnitude. In addition, we note that the central region has a larger thickness, as the fringes are no longer visible.  

\section{Discussion and conclusions}
In this paper we analyzed Marangoni spreading due to the injection of an IPA-water mixture on a water film. Using a microdrop generator, we were able to supply the IPA in a very localized area and at controlled flow rates, allowing for a detailed quantitative study of the spreading. The experimentally observed dynamics was explained through a balance between Marangoni and viscous stresses, and we proposed scaling laws, similar to those derived for insoluble surfactants \citep{Jensen1994}, that were verified in detail by a data-collapse. 

A key ingredient of the dynamics is that the surface tension gradient is spread out over the entire radius of the opening hole, i.e. $d\gamma/dr \sim \Delta \gamma/r$. This reduces the Marangoni driving force as the radius grows, which is why the spreading velocity decreases during the course of the experiment. This slowing down is markedly different from the ``delayed coalescence", observed when two sessile drops of miscible liquids come into contact \citep{Karpitschka2010, Borcia2011, Karpitschka2012}. Rather than coalescing, the drop of the larger surface tension was found to pull the other drop at constant velocity over the substrate. Such a constant velocity suggests that the surface tension gradient remains constant and localized where the two fluids meet, at least during the initial stages of the experiment \citep{Karpitschka2012}. The decrease in gradient observed in the present paper is much more reminiscent of the surfactant spreading \citep{Jensen1995, Hamraoui2004, Shearer2011, Fallest2010}, where similar scaling laws were derived, thus emphasizing the key importance of the flow geometry on Marangoni flows. 

A priori it is not obvious wether the IPA-water drops can be considered as insoluble surfactant. For this, one would need to know the detailed spatial distribution of the IPA concentration in water, both in the radial direction and across the depth of the liquid film. It is clear that the influence of diffusion needs to be considered for this. The duration of a typical experiment is about $0.1$s and the diffusion constant for IPA in water is $D \sim 10^{-9}\mbox{m}^2$/s, which leads to a diffusion length of $L=10\mu$m. This length is indeed totally negligible with respect to the radial scale, but only a bit larger than the characteristic thickness of the film. Hence, a lubrication approach for which the IPA is assumed to average quickly over the depth is not perfectly justified, but also the IPA does not remain perfectly at the interface. Similar to recent work on surfactants \citep{Swanson2013}, it would be interesting to further explore experimental possibilities to obtain more detailed information on the IPA distribution and the implications for the resulting Marangoni flow.

\appendix

\section{Calibrations}

We use the following cleaning procedure to give hydrophilic properties to a silica glass slide ($71m$m$\times71m$m). A precleaned glass slide is rinsed with pure water (resistivity ${\cal R}=18.2M\Omega$), wiped with an aqueous solution of a soap ($1\%$aq Hellmanex III), rinsed with pure water, then with isopropanol and dried with nitrogen. Afterwards, the substrate is placed inside a plasma cleaner (Harrick Plasma PDC-002) for $3$ minutes at $1.8m$bar. The hydrophilicity of this process is tested measuring the contact angle of a sessile drop of water.

To create the uniform water layer on the film, we fix the substrate on a spin coater with a vacuum pump. This is shown in Fig.~\ref{fig1}(a). Then, we continuously supply water to the substrate while spinning during $10$s. Afterwards, the water supply stops and the rotation continue during $10$s more. The resulting thickness is measured as a function of the spin rate (from $200$rpm to $600$rpm). The thickness is measured using a High-Resolution spectrometer (Ocean optics HR4000), similar to \cite{SpectroRef01} and \cite{Seiwert2011}.

We use a microdrop dispenser system (Microdrop Technologies MD-E-3000) to deposit a continuous supply of an IPA-water mixture on top of the water layer. The microdrop dispenser produces a continuous train of drops at frequencies varying from $1$Hz to $6000$Hz. For each IPA-water mixture we calibrate the settings of the microdrop. This calibration consists of finding microdrop settings that produce a monodisperse train of drops and measure the typical drop diameter, using the technique presented by \cite{vanderBos2011}, see Fig.~\ref{fig5}(b). This measurement is performed synchronizing a camera with the piezo pulse and a laser. The camera takes a picture at each cycle with a successive controlled delay. Different pictures at different moments of time allow to reconstruct the evolution of the stream of drops leaving the microdrop nozzle. The series of pictures are used to measure the drop diameter. The exact flow rate is calibrated by measuring the droplet size for each frequency. In order to achieve mono disperse size drops we do trial and error tests under different conditions.

We are grateful to I. Cantat, B. Dollet, S. Karpitschka, D. Lohse and H. Riegler for many interesting discussions. We also thank M-J. van der Meulen and E. Sandoval-Nava for their help in the microdrop device calibration. This work is sponsored by Lam Research, STW and NWO
by VIDI Grant No. 11304.

\bibliographystyle{plainnat}
\bibliography{bibliography}

\begin{thebibliography}{28}
\providecommand{\natexlab}[1]{#1}
\providecommand{\url}[1]{\texttt{#1}}
\expandafter\ifx\csname urlstyle\endcsname\relax
  \providecommand{\doi}[1]{doi: #1}\else
  \providecommand{\doi}{doi: \begingroup \urlstyle{rm}\Url}\fi

\bibitem[Afsar-Siddiqui et~al.(2003{\natexlab{a}})Afsar-Siddiqui, Luckham, and
  Matar]{Afsar_PART1_2003}
Abia~B Afsar-Siddiqui, Paul~F Luckham, and Omar~K Matar.
\newblock Unstable spreading of aqueous anionic surfactant solutions on liquid
  films. part 1. sparingly soluble surfactant.
\newblock \emph{Langmuir}, 19\penalty0 (3):\penalty0 696--702,
  2003{\natexlab{a}}.

\bibitem[Afsar-Siddiqui et~al.(2003{\natexlab{b}})Afsar-Siddiqui, Luckham, and
  Matar]{Afsar_PART2_2003}
Abia~B. Afsar-Siddiqui, Paul~F. Luckham, and Omar~K. Matar.
\newblock Unstable spreading of aqueous anionic surfactant solutions on liquid
  films. 2. highly soluble surfactant.
\newblock \emph{Langmuir}, 19\penalty0 (3):\penalty0 703--708,
  2003{\natexlab{b}}.

\bibitem[Borcia et~al.(2011)Borcia, Menzel, Bestehorn, Karpitschka, and
  Riegler]{Borcia2011}
R.~Borcia, S.~Menzel, M.~Bestehorn, S.~Karpitschka, and H.~Riegler.
\newblock Delayed coalescence of droplets with miscible liquids: Lubrication
  and phase field theories.
\newblock \emph{The European Physical Journal E}, 34\penalty0 (3):\penalty0
  1--9, 2011.
\newblock ISSN 1292-8941.
\newblock \doi{10.1140/epje/i2011-11024-9}.
\newblock URL \url{http://dx.doi.org/10.1140/epje/i2011-11024-9}.

\bibitem[{Borgas} and {Grotberg}(1988)]{Borgas1988}
M.~S. {Borgas} and J.~B. {Grotberg}.
\newblock Monolayer flow on a thin film.
\newblock \emph{Journal of Fluid Mechanics}, 193:\penalty0 151--170, Aug 1988.

\bibitem[Brochard-Wyart and de~Gennes(1992)]{DeGennes1992}
F.~Brochard-Wyart and P.G. de~Gennes.
\newblock Dynamics of partial wetting.
\newblock \emph{Advances in Colloid and Interface Science}, 39\penalty0
  (0):\penalty0 1 -- 11, 1992.
\newblock ISSN 0001-8686.

\bibitem[Fallest et~al.(2010)Fallest, Lichtenberger, Fox, and
  Daniels]{Fallest2010}
David~W Fallest, Adele~M Lichtenberger, Christopher~J Fox, and Karen~E Daniels.
\newblock Fluorescent visualization of a spreading surfactant.
\newblock \emph{New Journal of Physics}, 12\penalty0 (7):\penalty0 073029,
  2010.

\bibitem[{Gaver} and {Grotberg}(1990)]{Gaver1990}
D.~P. {Gaver} and J.~B. {Grotberg}.
\newblock The dynamics of a localized surfactant on a thin film.
\newblock \emph{Journal of Fluid Mechanics}, 213:\penalty0 127--148, apr 1990.

\bibitem[Hamraoui et~al.(2004)Hamraoui, Cachile, Poulard, and
  Cazabat]{Hamraoui2004}
A.~Hamraoui, M.~Cachile, C.~Poulard, and A.M. Cazabat.
\newblock Fingering phenomena during spreading of surfactant solutions.
\newblock \emph{Colloids and Surfaces A: Physicochemical and Engineering
  Aspects}, 250\penalty0 (1–3):\penalty0 215 -- 221, 2004.
\newblock ISSN 0927-7757.

\bibitem[Hosoi and Bush(2001)]{Hosoi2001}
AE~Hosoi and John~WM Bush.
\newblock Evaporative instabilities in climbing films.
\newblock \emph{Journal of Fluid Mechanics}, 442:\penalty0 217--239, 2001.

\bibitem[Jensen(1994)]{Jensen1994}
O.~E. Jensen.
\newblock Self-similar, surfactant-driven flows.
\newblock \emph{Physics of Fluids}, 6\penalty0 (3):\penalty0 1084--1094, 1994.

\bibitem[Jensen(1995)]{Jensen1995}
O.~E. Jensen.
\newblock The spreading of insoluble surfactant at the free surface of a deep
  fluid layer.
\newblock \emph{Journal of Fluid Mechanics}, 293:\penalty0 349--378, 6 1995.
\newblock ISSN 1469-7645.

\bibitem[Karpitschka and Riegler(2010)]{Karpitschka2010}
Stefan Karpitschka and Hans Riegler.
\newblock Quantitative experimental study on the transition between fast and
  delayed coalescence of sessile droplets with different but completely
  miscible liquids.
\newblock \emph{Langmuir}, 26\penalty0 (14):\penalty0 11823--11829, 2010.

\bibitem[Karpitschka and Riegler(2012)]{Karpitschka2012}
Stefan Karpitschka and Hans Riegler.
\newblock Noncoalescence of sessile drops from different but miscible liquids:
  Hydrodynamic analysis of the twin drop contour as a self-stabilizing
  traveling wave.
\newblock \emph{Phys. Rev. Lett.}, 109:\penalty0 066103, Aug 2012.
\newblock \doi{10.1103/PhysRevLett.109.066103}.
\newblock URL \url{http://link.aps.org/doi/10.1103/PhysRevLett.109.066103}.

\bibitem[Lee and Starov(2009)]{Lee2009}
KS~Lee and VM~Starov.
\newblock Spreading of surfactant solutions over thin aqueous layers at low
  concentrations: Influence of solubility.
\newblock \emph{Journal of colloid and interface science}, 329\penalty0
  (2):\penalty0 361--365, 2009.

\bibitem[Leenaars et~al.(1990)Leenaars, Huethorst, and Van~Oekel]{Leenaars1990}
A.~F.~M. Leenaars, J.~A.~M. Huethorst, and J.~J. Van~Oekel.
\newblock Marangoni drying: A new extremely clean drying process.
\newblock \emph{Langmuir}, 6\penalty0 (11):\penalty0 1701--1703, 1990.

\bibitem[Marra and Huethorst(1991)]{Marra1991}
J.~Marra and J.~A.~M. Huethorst.
\newblock Physical principles of marangoni drying.
\newblock \emph{Langmuir}, 7\penalty0 (11):\penalty0 2748--2755, 1991.

\bibitem[Matar and Craster(2001)]{Matar2001}
O.~K. Matar and R.~V. Craster.
\newblock Models for marangoni drying.
\newblock \emph{Physics of Fluids}, 13\penalty0 (7):\penalty0 1869--1883, 2001.

\bibitem[Matar and Craster(2009)]{matar2009dynamics}
OK~Matar and RV~Craster.
\newblock Dynamics of surfactant-assisted spreading.
\newblock \emph{Soft Matter}, 5\penalty0 (20):\penalty0 3801--3809, 2009.

\bibitem[Matar and Troian(1999)]{Matar1999}
Omar~K. Matar and Sandra~M. Troian.
\newblock Spreading of a surfactant monolayer on a thin liquid film: Onset and
  evolution of digitated structures.
\newblock \emph{Chaos: An Interdisciplinary Journal of Nonlinear Science},
  9\penalty0 (1):\penalty0 141--153, 1999.

\bibitem[Ngo et~al.(2013)Ngo, Yu, and Lin]{ViscoProp01}
Trung~Truc Ngo, T.~Leon Yu, and Hsiu-Li Lin.
\newblock Influence of the composition of isopropyl alcohol/water mixture
  solvents in catalyst ink solutions on proton exchange membrane fuel cell
  performance.
\newblock \emph{Journal of Power Sources}, 225\penalty0 (0):\penalty0 293 --
  303, 2013.
\newblock ISSN 0378-7753.

\bibitem[O'Brien(1993)]{OBrien1993}
S.~B. G.~M. O'Brien.
\newblock On marangoni drying: nonlinear kinematic waves in a thin film.
\newblock \emph{Journal of Fluid Mechanics}, 254:\penalty0 649--670, 9 1993.

\bibitem[Peterson and Shearer(2011)]{Shearer2011}
Ellen~R. Peterson and Michael Shearer.
\newblock Radial spreading of a surfactant on a thin liquid film.
\newblock \emph{Applied Mathematics Research eXpress}, 2011\penalty0
  (1):\penalty0 1--22, 2011.

\bibitem[Seiwert et~al.(2011)Seiwert, Clanet, and Qu\'er\'e]{Seiwert2011}
Jacopo Seiwert, Christophe Clanet, and David Qu\'er\'e.
\newblock Coating of a textured solid.
\newblock \emph{Journal of Fluid Mechanics}, 669:\penalty0 55--63, 2 2011.
\newblock ISSN 1469-7645.

\bibitem[Snoeijer et~al.(2008)Snoeijer, Ziegler, Andreotti, Fermigier, and
  Eggers]{SpectroRef01}
J.~H. Snoeijer, J.~Ziegler, B.~Andreotti, M.~Fermigier, and J.~Eggers.
\newblock Thick films of viscous fluid coating a plate withdrawn from a liquid
  reservoir.
\newblock \emph{Phys. Rev. Lett.}, 100:\penalty0 244502, Jun 2008.

\bibitem[Swanson et~al.(2013)Swanson, Strickland, Shearer, and
  Daniels]{Swanson2013}
Ellen~R Swanson, Stephen~L Strickland, Michael Shearer, and Karen~E Daniels.
\newblock Surfactant spreading on a thin liquid film: Reconciling models and
  experiments.
\newblock \emph{arXiv preprint arXiv:1306.4881}, 2013.

\bibitem[van~der Bos et~al.(2011)van~der Bos, Zijlstra, Gelderblom, and
  Versluis]{vanderBos2011}
Arjan van~der Bos, Aaldert Zijlstra, Erik Gelderblom, and Michel Versluis.
\newblock ilif: illumination by laser-induced fluorescence for single flash
  imaging on a nanoseconds timescale.
\newblock \emph{Experiments in Fluids}, 51\penalty0 (5):\penalty0 1283--1289,
  2011.
\newblock ISSN 0723-4864.
\newblock \doi{10.1007/s00348-011-1146-7}.

\bibitem[van~der Veen et~al.(2012)van~der Veen, Tran, Lohse, and
  Sun]{Roeland2012}
Roeland C.~A. van~der Veen, Tuan Tran, Detlef Lohse, and Chao Sun.
\newblock Direct measurements of air layer profiles under impacting droplets
  using high-speed color interferometry.
\newblock \emph{Phys. Rev. E}, 85:\penalty0 026315, Feb 2012.

\bibitem[Warner et~al.(2004)Warner, Craster, and Matar]{Matar2004}
MRE Warner, RV~Craster, and OK~Matar.
\newblock Fingering phenomena associated with insoluble surfactant spreading on
  thin liquid films.
\newblock \emph{Journal of Fluid Mechanics}, 510:\penalty0 169--200, 2004.

\end{thebibliography}

\end{document}